\documentclass[sigconf]{acmart}

\AtBeginDocument{%
  }

\usepackage{multirow}
\usepackage{multirow}
\usepackage{enumitem}
\usepackage{xspace}
\usepackage{cleveref}
\usepackage{subfig}
\usepackage{soul}
\usepackage{pgfplots, pgfplotstable}
\usepackage{booktabs}   
\usepackage{pifont}
\usepackage{makecell}
\usepackage{multirow}            
\usepackage{amsfonts}            
\usepackage{graphicx}            
\usepackage{duckuments} 
\usepackage[ruled,vlined]{algorithm2e}

\pgfplotsset{compat=1.17}
\usepgfplotslibrary{colorbrewer} 

\definecolor{gray}{rgb}{0.83, 0.83, 0.83}

\usepackage{tcolorbox}

\newcommand{\tikzcircle}[2][red,fill=red]{\tikz[baseline=-0.5ex]\draw[#1,radius=#2] (0,0) circle ;}

\definecolor{sisinflab}{HTML}{5a9699}

\usepackage{geometry}
\usepackage{array}
\usepackage{xcolor}

\usepackage{color, colortbl}
	
\definecolor{Gray}{gray}{0.9}

\setcopyright{acmlicensed}
\copyrightyear{2024}
\acmYear{2024}
\acmDOI{XXXXXXX.XXXXXXX}

\acmBooktitle{Eighteenth ACM Conference on Recommender Systems (RecSys '24), October 14--18, 2024, Bari, Italy}
\acmISBN{978-1-4503-XXXX-X/24/10}

\author{Daniele Malitesta}
\authornote{Corresponding authors.}
\authornote{Work done while at Politecnico di Bari as a PhD student.}
\orcid{0000-0003-2228-0333}
\affiliation{%
\institution{Université Paris-Saclay CentraleSupélec, Inria}
  \city{Gif-sur-Yvette}
  \country{France}}
\email{daniele.malitesta@centralesupelec.fr}

\author{Claudio Pomo}
\authornotemark[1]
\orcid{0000-0001-5206-3909}
\affiliation{%
\institution{Politecnico di Bari}
  \city{Bari}
  \country{Italy}}
\email{claudio.pomo@poliba.it}

\author{Vito Walter Anelli}
\orcid{0000-0002-5567-4307}
\affiliation{%
\institution{Politecnico di Bari}
  \city{Bari}
  \country{Italy}}
\email{vitowalter.anelli@poliba.it}

\author{Alberto Carlo Maria Mancino}
\orcid{0000-0001-8027-9475}
\affiliation{%
\institution{Politecnico di Bari}
  \city{Bari}
  \country{Italy}}
\email{alberto.mancino@poliba.it}

\author{Tommaso {Di Noia}}
\orcid{0000-0002-0939-5462}
\affiliation{%
\institution{Politecnico di Bari}
  \city{Bari}
  \country{Italy}}
\email{tommaso.dinoia@poliba.it}

\author{Eugenio {Di Sciascio}}
\orcid{0000-0002-5484-9945}
\affiliation{%
\institution{Politecnico di Bari}
  \city{Bari}
  \country{Italy}}
\email{eugenio.disciascio@poliba.it}


\begin{document}

\title{A Novel Evaluation Perspective on GNNs-based Recommender Systems through the Topology of the User-Item Graph}


\renewcommand{\shortauthors}{Malitesta, Pomo, Anelli, Mancino, Di Noia, \& Di Sciascio}

\begin{abstract}

Recently, graph neural networks (GNNs)-based recommender systems have encountered great success in recommendation. As the number of GNNs approaches rises, some works have started questioning the theoretical and empirical reasons behind their superior performance. Nevertheless, this investigation still disregards that GNNs treat the recommendation data as a topological graph structure. Building on this assumption, in this work, we provide a novel evaluation perspective on GNNs-based recommendation, which investigates the impact of the graph topology on the recommendation performance. To this end, we select some (topological) properties of the recommendation data and three GNNs-based recommender systems (i.e., LightGCN, DGCF, and SVD-GCN). Then, starting from three popular recommendation datasets (i.e., Yelp2018, Gowalla, and Amazon-Book) we sample them to obtain 1,800 size-reduced datasets that still resemble the original ones but can encompass a wider range of topological structures. We use this procedure to build a large pool of samples for which data characteristics and recommendation performance of the selected GNNs models are measured. Through an explanatory framework, we find strong correspondences between graph topology and GNNs performance, offering a novel evaluation perspective on these models.

\end{abstract}




\copyrightyear{2024}
\acmYear{2024}
\setcopyright{rightsretained}
\acmConference[RecSys '24]{18th ACM Conference on Recommender
Systems}{October 14--18, 2024}{Bari, Italy}
\acmBooktitle{18th ACM Conference on Recommender Systems (RecSys '24),
October 14--18, 2024, Bari, Italy}\acmDOI{10.1145/3640457.3688070}
\acmISBN{979-8-4007-0505-2/24/10}

\renewcommand\footnotetextcopyrightpermission[1]{}

\maketitle

\section{Introduction and Motivations}

Graph neural networks (GNNs)-based recommender systems~\cite{DBLP:journals/csur/WuSZXC23, DBLP:journals/corr/abs-2310-11270} are regarded as well-established approaches within the plethora of existing recommendation algorithms, together with traditional model families encompassing, among others, neighborhood-based solutions~\cite{DBLP:conf/www/SarwarKKR01, DBLP:journals/tiis/PaudelCNB17}, latent-factor approaches~\cite{DBLP:conf/uai/RendleFGS09, DBLP:conf/icdm/Rendle10, DBLP:conf/recsys/AnelliNSFM21, DBLP:journals/tors/FerraraAMNS23}, as well as more recent frameworks based, for instance, on deep neural networks~\cite{DBLP:conf/recsys/CovingtonAS16, DBLP:conf/www/HeLZNHC17}, variational autoencoders~\cite{DBLP:conf/sigir/AskariSS21, DBLP:conf/wsdm/ShenbinATMN20}, and generative models~\cite{DBLP:conf/sigir/WangYZGXWZZ17, DBLP:conf/sigir/0001HDC18}. 

In early years, GNNs-based recommender systems proposed to design and apply GNN message-passing layers from the graph representation learning domain (i.e., graph convolutional layers such as GraphSAGE~\cite{DBLP:conf/nips/HamiltonYL17} and GCN~\cite{DBLP:conf/iclr/KipfW17}) to tailor them for the recommendation task. In this respect, pioneer works such as~\cite{DBLP:journals/corr/BergKW17, DBLP:conf/kdd/YingHCEHL18, DBLP:conf/sigir/Wang0WFC19} put the selected GNN layer on top of the traditional latent-factor recommendation module; that is, users' and items' embeddings are initially refined through the message-passing procedure that can catch existing user-item relationships at multiple hops. 

Over the years, other solutions tried to improve the existing techniques in various manners, inspired by the advances in graph representation learning. Among them, some approaches proposed to simplify~\cite{DBLP:conf/icml/WuSZFYW19} the message-passing procedure through the removal of feature transformation and non-linearities~\cite{DBLP:conf/sigir/0001DWLZ020, DBLP:conf/aaai/ChenWHZW20}, while others exploited the graph edges to weight their importance~\cite{DBLP:conf/cikm/AnelliDNSFMP22, DBLP:conf/recsys/MancinoFBMNS23}, in some cases through attention weights (as in the original GAT~\cite{DBLP:conf/iclr/VelickovicCCRLB18} architecture) to prune noisy interactions and disentangle them at the granularity of intents~\cite{DBLP:conf/icml/Ma0KW019, DBLP:conf/sigir/ZhangL0WSLSZDZ22, DBLP:conf/sigir/WangJZ0XC20, DBLP:conf/www/WangHWYL0C21}.

More recently, novel trends involved semi-supervised~\cite{DBLP:conf/cikm/HuangXW0Y22} and contrastive learning~\cite{DBLP:conf/nips/KhoslaTWSTIMLK20} to tackle the data sparsity of the user-item graph through ad-hoc graph augmentations~\cite{DBLP:conf/www/LinTHZ22, DBLP:conf/iclr/Cai0XR23, DBLP:conf/sigir/0001OM22, DBLP:conf/sigir/YuY00CN22, DBLP:conf/sigir/WuWF0CLX21, DBLP:journals/tkde/YuXCCHY24}. In parallel, other techniques suggested that addressing foundational problems in graph representation learning (such as over-smoothing~\cite{DBLP:conf/aaai/ChenLLLZS20}, over-squashing~\cite{DBLP:conf/iclr/ToppingGC0B22}, or both phenomena~\cite{DBLP:conf/cikm/GiraldoSBM23}) could also improve GNNs-based recommendation by surpassing the traditional concept of message-passing~\cite{DBLP:conf/cikm/MaoZXLWH21, DBLP:conf/cikm/ShenWZSZLL21, DBLP:conf/cikm/PengSM22, DBLP:journals/corr/abs-2403-17416}. 

As new GNN solutions are continuously proposed in the literature, a fundamental question arises: ``\textit{\textbf{Why GNNs-based recommendation works so well?}}''. Indeed, the dilemma is pertinent since more and more techniques are introduced daily, and knowing which strategies may better suit the recommendation problem (and why) has become imperative. In this respect, the literature recognizes that two complementary research directions have emerged so far. 

On the one side, some works investigated the \textbf{\textit{theoretical}} reasons behind the performance of GNNs-based recommender systems. Among others,~\citet{DBLP:conf/cikm/ShenWZSZLL21} observed that the graph convolution operation could be behind the success of GNNs-based recommendation, working as a smoothing filter on nodes features; by leveraging the theory of graph signal processing, the authors revisited several latent factors and GNNs-based models under a shared framework built upon the smoothing filtering operation, eventually proposing a lightweight closed-form solution that can reach the performance of other less shallow recommendation solutions. More recently,~\citet{DBLP:journals/corr/abs-2403-17416} discussed the role of over-smoothing and over-correlation in GNNs-based recommendation; by acknowledging that the two phenomena are intrinsically bound, the authors proposed an adaptive de-correlation loss function to tackle the two issues at once.

On the other side, some works explored the \textbf{\textit{experimental}} reasons behind the performance of GNNs-based recommender systems. For instance, the work by~\citet{DBLP:conf/wsdm/WangZS22} tried to generalize the GNN recommendation pipeline into separate building blocks that can be changed and removed accordingly; by exploring the numerous different architectures in the literature, the authors defined a search space that could easily be adapted for every recommendation scenario. Then, the works in~\cite{DBLP:conf/ecir/AnelliDNMPP23, DBLP:conf/recsys/AnelliDNSFMP22} provided an outlook on accuracy and beyond-accuracy recommendation performance of GNNs-based recommender systems, recognizing node representation and neighborhood exploration as the key strategy patterns to build any recommendation model based upon message-passing. More recently, the work by~\citet{DBLP:conf/recsys/AnelliMPBSN23} proposed a reproducibility analysis on popular GNNs-based recommender systems spanning several existing solutions; while confirming the effective reproducibility of such models for the experimental settings presented in the original papers, the authors outlined how the commonly observed performance may drastically change on usually untested settings (i.e, other baselines or datasets), further suggesting that dataset intrinsic properties (such as the average node degree) may be greatly influencing the final performance. 

Nevertheless, we believe both outlined research directions still disregard one simple (but potentially fundamental) aspect. Since GNNs are designed to treat the recommendation data as a bipartite and undirected graph, there is a whole set of \textbf{\textit{topological structures}} of the user-item graph that such models should be able to recognize and exploit to boost the recommendation performance. Thus, answering the question: ``\textit{\textbf{Why GNNs-based recommendation works so well?}}'', would eventually imply answering another question: ``\textit{\textbf{Which topological dataset characteristics are GNNs-based recommender systems able to capture? Are those properties influencing the recommendation performance?}}''.

Noticeably, our intuition finds supporting motivations in the recent literature. On the one hand, works in graph representation learning have started questioning what is the role of the graph structure (both topology and node features) in the performance of GNNs models~\cite{DBLP:conf/icml/WuSZFYW19, DBLP:conf/www/WeiZH22, DBLP:conf/www/0002ZPNG0CH22, DBLP:conf/iclr/KlicperaBG19, DBLP:conf/iclr/ZhaoA20, DBLP:conf/icml/Abu-El-HaijaPKA19, DBLP:conf/iclr/KlicperaBG19, DBLP:conf/nips/LuanZCP19, DBLP:conf/icml/XuLTSKJ18, DBLP:conf/iclr/RongHXH20, DBLP:journals/corr/abs-2308-09570}; for instance, it is currently acknowledged that GNNs excel in node classification tasks when working on graphs exhibiting high levels of \textit{homophily}~\cite{DBLP:conf/nips/ZhuYZHAK20}; similarly, recent work suggested that node degree can greatly influence the performance of GNNs in both node classification and link prediction tasks~\cite{DBLP:journals/corr/abs-2310-04612}. On the other hand, existing studies in recommendation outlined that dataset characteristics (such as the dataset sparsity) may impact the final performance 
~\cite{DBLP:journals/tmis/AdomaviciusZ12, DBLP:conf/sigir/DeldjooNSM20, DBLP:journals/corr/abs-2308-10778}. 

\subsection{Our contributions}

For all the outlined aspects, in this work, we propose a novel evaluation perspective on GNNs-based recommender systems, that accounts for the \textbf{\textit{topological properties}} of the user-item graph. While our paper ideally settles in between the \textit{\textbf{theoretical}} and \textbf{\textit{experimental}} research lines outlined above, a fundamental starting basis is provided by the works in~\cite{DBLP:conf/recsys/AnelliMPBSN23, DBLP:conf/um/MalitestaPANF23}, where the authors eventually discuss how node degree information (propagated at multiple hops) may affect recommendation accuracy for different users' groups in the system. We decide to extend their reproducibility and experimental settings by proposing a \textbf{\textit{novel evaluation pipeline}} for GNNs-based recommender systems, where we seek to find statistical correspondences between the topological properties of the user-item graph and the recommendation performance of such models. 
Thus, our contributions are summarized as:

\begin{enumerate}
    \item We select five \textit{classical} dataset characteristics (presented in~\cite{DBLP:journals/tmis/AdomaviciusZ12, DBLP:conf/sigir/DeldjooNSM20}) and three additional \textit{topological} dataset characteristics~\cite{DBLP:journals/socnet/LatapyMV08, PhysRevE.67.026126} (node degree, clustering coefficient, and degree assortativity); for the \textit{topological} ones, we provide a re-interpretation under the lens of recommendation.
    \item We consider three popular and recent GNNs-based recommendation approaches (i.e., LightGCN~\cite{DBLP:conf/sigir/0001DWLZ020}, DGCF~\cite{DBLP:conf/sigir/WangJZ0XC20}, and SVD-GCN~\cite{DBLP:conf/cikm/PengSM22}) and work on their formulations to explicitly highlight the presence of the \textit{topological} properties.
    \item We propose a novel evaluation pipeline that investigates the influence of dataset characteristics on the recommendation performance of GNNs-based recommender systems; this involves the adoption of ad-hoc graph sampling strategies 
    to generate a large suite of 1,800 small datasets that resemble three popular recommendation datasets (i.e., Yelp2018, Gowalla, and Amazon-Book) but encompass a wider range of topological structures.
    \item Through an explanatory framework, we show that strong correspondences exist between dataset characteristics and recommendation performance for GNNs-based models, offering a novel evaluation perspective on such approaches.
    \item To assess the goodness of the evaluation pipeline, we analyze the possible impact of node- and edge-dropout on the generated explanations, proving that their joint adoption is beneficial to the statistical significance of the explanatory framework.
\end{enumerate}
 
To foster reproducibility, we release the code to replicate our evaluation pipeline: \url{https://github.com/sisinflab/Topology-Graph-Collaborative-Filtering}. 
\section{Topological data characteristics} \label{sec:topological_char}
In this section, we provide a formal description of the recommendation dataset in terms of its \textit{topological} characteristics~\cite{DBLP:journals/socnet/LatapyMV08, PhysRevE.67.026126}, treating the data as a bipartite and undirected graph; noticeably, for each topological property, we also offer a re-interpretation under the lens of recommendation. As our analysis will eventually assess also the influence of \textit{classical} recommendation dataset characteristics already presented in~\cite{DBLP:journals/tmis/AdomaviciusZ12, DBLP:conf/sigir/DeldjooNSM20}, and for the sake of space, we suggest the reader refer to those works for a formal presentation.  








In a recommendation system, we denote with $\mathcal{U}$ and $\mathcal{I}$ the sets of users and items, respectively, where $|\mathcal{U}| = U$ and $|\mathcal{I}| = I$. Then, we indicate with $\mathbf{R} \in \mathbb{R}^{U \times I}$ the interaction matrix collecting user-item interactions in the form of implicit feedback (i.e., $\mathbf{R}_{u, i} = 1$ if user $u \in \mathcal{U}$ interacted with item $i \in \mathcal{I}$, 0 otherwise). 

The interaction matrix $\mathbf{R}$ is leveraged to define the adjacency matrix $\mathbf{A} \in \mathbb{R}^{(U + I) \times {(U + I)}}$ representing the bidirectional interactions between users and items:
\begin{equation}
    \mathbf{A} = \begin{bmatrix}
    0 & \mathbf{R} \\
    \mathbf{R}^\top & 0 
    \end{bmatrix},
\end{equation}
leading to $\mathcal{G} = \{\mathcal{U} \cup \mathcal{I}, \mathbf{A}\}$ as the user-item bipartite and undirected graph. Moreover, we connote the user- and item-\textit{projected} graphs as $\mathcal{G}_\mathcal{U} = \{\mathcal{U}, \mathbf{A}^{\mathcal{U}}\}$ and $\mathcal{G}_\mathcal{I} = \{\mathcal{I}, \mathbf{A}^{\mathcal{I}}\}$. In this respect, let $\mathbf{R}^{\mathcal{U}}$ and $\mathbf{R}^{\mathcal{I}}$ be the user-user and item-item interaction matrices:
\begin{equation}
    \mathbf{R}^{\mathcal{U}} = \mathbf{R} \cdot \mathbf{R}^\top, \qquad \mathbf{R}^{\mathcal{I}} = \mathbf{R}^\top \cdot \mathbf{R},
\end{equation}
which indicate the co-occurrences among users and items, respectively. Trivially, the corresponding adjacency matrices $\mathbf{A}^{\mathcal{U}}$ and $\mathbf{A}^{\mathcal{I}}$ are obtained as:
\begin{equation}
    \mathbf{A}^{\mathcal{U}} = \mathbf{R}^{\mathcal{U}}, \qquad \mathbf{A}^{\mathcal{I}} = \mathbf{R}^{\mathcal{I}}.
\end{equation}
We use the introduced concepts and notations to describe three topological aspects of the user-item graph and re-interpret them under the lens of recommender systems.

\textbf{\ul{Node degree}.} Let $\mathcal{N}_u = \{i\;|\; \mathbf{R}_{u, i} = 1\}$ and $\mathcal{N}_i = \{u\;|\;\mathbf{R}_{u, i} = 1\}$ be the neighborhood sets for $u$ and $i$, respectively. By generalizing this definition, let $\mathcal{N}^{(l)}_u$ and $\mathcal{N}^{(l)}_i$ be the sets of neighborhood nodes for user $u$ and item $i$ at $l$ distance hops. Thus, the node degrees for $u$ and $i$ (i.e., $\sigma_u = |\mathcal{N}^{(1)}_u|$ and $\sigma_i = |\mathcal{N}^{(1)}_i|$) represent the number of item and user nodes directly connected with $u$ and $i$, respectively. The average user and item node degrees are:
\begin{equation}
    \sigma_\mathcal{U} = \frac{1}{U} \sum_{u \in \mathcal{U}} |\mathcal{N}^{(1)}_u|, \qquad \sigma_\mathcal{I} = \frac{1}{I} \sum_{i \in \mathcal{I}} |\mathcal{N}^{(1)}_i|.
\end{equation}

\vspace{1em}
\noindent\textsc{\bfseries RecSys re-interpretation.} \textit{The node degree in the user-item graph stands for the number of items (users) interacted by a user (item). This is related to the cold-start issue in recommendation, where cold users denote low activity on the platform, while cold items are niche products.}
\vspace{1em}

Node degree alone still fails to provide a deeper outlook on the user-item graph. The following topological characteristics, derived from node degree, expand its formulation to other viewpoints.

\textbf{\ul{Clustering coefficient}.} For each partition in a bipartite graph, it is interesting to recognize clusters of nodes in terms of how their neighborhoods overlap, independently of the respective sizes. Let $v$ and $w$ be two nodes from the same partition (e.g., user nodes). Their similarity is the intersection over union of their neighborhoods~\cite{DBLP:journals/socnet/LatapyMV08}. By evaluating the metric node-wise, we obtain:
\begin{equation}
    \gamma_{v} = \frac{\sum_{w \in \mathcal{N}^{(2)}_v}\gamma_{v, w}}{|\mathcal{N}_v^{(2)}|}, \qquad \text{with} \quad  \gamma_{v, w} = \frac{|\mathcal{N}^{(1)}_{v} \cap \mathcal{N}^{(1)}_{w}|}{|\mathcal{N}^{(1)}_{v} \cup \mathcal{N}^{(1)}_{w}|},
\end{equation}
where $\mathcal{N}_v^{(2)}$ is the second-order neighborhood set of $v$. In this case, we leverage the second-order neighborhood because, in a bipartite graph, nodes from the same partition are connected at (multiple of) 2 hops. The average clustering coefficient on $\mathcal{U}$ and $\mathcal{I}$ is:
\begin{equation}
    \gamma_{\mathcal{U}} = \frac{1}{U} \sum_{u \in \mathcal{U}} \gamma_u, \qquad \gamma_{\mathcal{I}} = \frac{1}{I} \sum_{i \in \mathcal{I}} \gamma_i.
\end{equation}

\vspace{1em}
\noindent\textsc{\bfseries RecSys re-interpretation.} \textit{High values of the clustering coefficient indicate that there exists a substantial number of co-occurrences among nodes from the same partition. For instance, when considering the user-side formula, the average clustering coefficient increases if several users share most of their interacted items. The intuition aligns with the rationale behind collaborative filtering: two users are likely to show similar preferences when they interact with the same items.}
\vspace{1em}

The clustering coefficient allows the description of broader portions of the user-item graph compared to the semantics conveyed by node degree. Indeed, the measure takes nodes at 2 hops (i.e., user-item-user and item-user-item connections). Nevertheless, we may want to capture properties for even more extended regions of the graph. For this reason, we introduce one last topological characteristic that goes beyond the 2-hop distance among nodes.   

\textbf{\ul{Degree assortativity}.}
\label{sec:assortativity} In real-world graphs, nodes tend to gather when they share similar characteristics. Such a tendency is measured through the assortativity coefficient. Depending on the semantics of ``node similarity'', there exist different formulations for assortativity~\cite{PhysRevE.67.026126}. For the sake of this work, we consider the assortativity coefficient based on the scalar properties of graph nodes, for instance, their degree. Let $\mathcal{D} = \{d_1, d_2, \dots\}$ be the set of unique node degrees in the graph, and let $e_{d_h, d_k}$ be the fraction of edges connecting nodes with degrees $d_h$ and $d_k$. Then, let $q_{d_h}$ be the probability distribution to choose a node with degree $d_h$ after having selected a node with the same degree (i.e., the \textit{excess} degree distribution). The degree assortativity coefficient is calculated as:
\begin{equation}
    \rho = \frac{\sum\limits_{d_h, d_k}d_h d_k(e_{d_h, d_k} - q_{d_h} q_{d_k})}{std^2_q},
\end{equation}
where $std_q$ is the standard deviation of the distribution $q$. Note that, for its formulation, the degree assortativity is similar to a correlation measure (e.g., Pearson correlation). Following the same rationale of the clustering coefficient, we are interested in finding similarity patterns among nodes from the same partition. For this reason, we first apply the projection of the user-item bipartite graph for both users and items to obtain the user- (i.e., $\mathcal{G}_{\mathcal{U}}$) and item- (i.e., $\mathcal{G}_{\mathcal{I}}$) projected graphs. Then, we calculate the degree assortativity coefficients for $\mathcal{G}_{\mathcal{U}}$ and $\mathcal{G}_{\mathcal{I}}$, namely, $\rho_{\mathcal{U}}$ and $\rho_{\mathcal{I}}$.

\vspace{2em}
\noindent\textsc{\bfseries RecSys re-interpretation.} \textit{In recommendation, the degree assortativity calculated user- and item-wise represents the tendency of users with the same activity level on the platform and items with the same popularity to gather, respectively. Since we calculate the degree assortativity on the complete user-user and item-item co-occurrence graphs, we deem this characteristic to provide a broader view of the dataset than the clustering coefficient. For this reason, to give an intuition of degree assortativity, we borrow the concept of search space traversal depth in search algorithms theory. That is, we re-interpret degree assortativity in recommendation as a topological characteristic showing a strong look-ahead nature.}
\section{Topological Characteristics in Graph-based recommendation}
\label{sec:top-char-graph-collab}

Since GNNs-based recommender systems are designed to work on the bipartite and undirected user-item graph, we seek to understand how and to what extent such models (explicitly) integrate \textit{topological} properties into their formulations. Thus, we select three popular and recent approaches in GNNs-based recommendation (i.e., LightGCN~\cite{DBLP:conf/sigir/0001DWLZ020}, DGCF~\cite{DBLP:conf/sigir/WangJZ0XC20}, and SVD-GCN~\cite{DBLP:conf/cikm/PengSM22}). Then, we re-formulate their techniques to make the \textit{topological} data characteristics explicitly emerge. 

As additional background with respect to the above, we introduce the notations $\mathbf{e}_u \in \mathbb{R}^b$ and $\mathbf{e}_i \in \mathbb{R}^b$ as the initial embeddings of the nodes for user $u$ and item $i$, respectively, where $b << U, I$. Then, in the case of message-propagation at different layers, we also introduce the notations $\mathbf{e}_u^{(l)}$ and $\mathbf{e}_i^{(l)}$ to indicate the updated node embeddings for user $u$ and item $i$ after $l$ propagation layers, with $0 \leq l \leq L$ (note that $\mathbf{e}_u^{(0)} = \mathbf{e}_u$ and $\mathbf{e}_i^{(0)} = \mathbf{e}_i$).

\textbf{\ul{LightGCN}.} \citet{DBLP:conf/sigir/0001DWLZ020} propose to lighten the graph convolutional layer presented in~\citet{DBLP:conf/iclr/KipfW17} for the recommendation task. Specifically, their layer removes feature transformation and non-linearities:
\begin{equation}
\label{eq:lightgcn}
    \mathbf{e}_u^{(l)} = \sum\limits_{i' \in \mathcal{N}_u^{(1)}} \frac{A_{ui'} \mathbf{e}_{i'}^{(l - 1)}}{\sqrt{\sigma_u  \sigma_{i'}}}, \quad \mathbf{e}_i^{(l)} = \sum\limits_{u' \in \mathcal{N}_i^{(1)}} \frac{A_{iu'} \mathbf{e}_{u'}^{(l - 1)}}{\sqrt{\sigma_i \sigma_{u'}}},
\end{equation}
where each neighbor contribution is weighted through the corresponding entry in the normalized Laplacian adjacency matrix to flatten the differences among nodes with high and low degrees. Since $A_{ui'} = 1, \text{ } \forall i' \in \mathcal{N}^{(1)}_u$ (the dual holds for $A_{iu'}$), the contribution weighting comes only from the denominator.  

\textbf{\ul{DGCF}.} \citet{DBLP:conf/sigir/WangJZ0XC20} assume that user-item interactions are decomposed into a set of independent intents, representing the specific aspects users may be interested in when interacting with items. In this respect, the authors propose to iteratively learn a set of weighted adjacency matrices $\{\mathbf{\tilde{A}}_1, \mathbf{\tilde{A}}_2, \dots\}$, where each of them records the user-item importance weights based on the specific intent it represents. Then, they introduce a graph disentangling layer for each weighted adjacency matrix:
\begin{equation}
    \mathbf{e}_{u, *}^{(l)} = \sum\limits_{i' \in \mathcal{N}_u^{(1)}} \frac{\tilde{A}_{ui', *} \mathbf{e}_{i', *}^{(l - 1)}}{\sqrt{\sigma_{u, *} \sigma_{i', *}}}, \quad \mathbf{e}_{i, *}^{(l)} = \sum\limits_{u' \in \mathcal{N}_i^{(1)}} \frac{\tilde{A}_{iu', *} \mathbf{e}_{u', *}^{(l - 1)}}{\sqrt{\sigma_{i, *} \sigma_{u', *}}},
\end{equation}
where $\tilde{A}_{ui', *}$ and $\mathbf{e}_{i',*}^{(l-1)}$ are the learned importance weight of user $u$ on item $i'$ and the embedding of item $i'$ for any intent, while $\sigma_{u,*}$ is the corresponding node degree calculated on $\mathbf{\tilde{A}_{*}}$ (the same applies for the item side).

\textbf{\ul{SVD-GCN}.} \citet{DBLP:conf/cikm/PengSM22} propose a reformulation of the GCN-based message-passing which leverages the similarities between graph convolutional layers and singular value decomposition (i.e., SVD). Specifically, they rewrite the message-passing introduced in LightGCN by making two aspects explicitly emerge, namely: (i) even- and odd-connection message aggregations, and (ii) singular values and vectors obtained by decomposing the user-item interaction matrix $\mathbf{R}$ through SVD. On such basis, the authors' assumption is that the traditional graph convolutional layer intrinsically learns a low-rank representation of the user-item interaction matrix where components corresponding to larger singular values tend to be enhanced. 
They reinterpret the over-smoothing effect as an increasing gap between singular values when stacking more and more layers. 
The embeddings for users and items are obtained as follows:
\begin{equation}
    \mathbf{e}_u = \mathbf{p}_u exp(a_1 \boldsymbol\lambda) \cdot \mathbf{W}, \qquad
    \mathbf{e}_i = \mathbf{q}_i exp(a_1 \boldsymbol\lambda) \cdot \mathbf{W},
\end{equation}
where: (i) $\mathbf{p}_u$ and $\mathbf{q}_i$ are the left and right singular vectors of the normalized user-item interaction matrix for user $u$ and item $i$; (ii) $exp()$ is the exponential function; (iii) $a_1$ is a tunable hyper-parameter of the model; (iv) $\boldsymbol\lambda$ is the vector of the largest singular values of the normalized user-item matrix; (v) $\mathbf{W}$ is a trainable matrix to perform feature transformation. Note that the highest singular value $\lambda_{max}$ and the maximum node degree $\max(\mathcal{D})$ in the user-item interaction matrix are associated by the following inequality:
\begin{equation}
    \lambda_{max} \leq \frac{\max(\mathcal{D})}{\max(\mathcal{D}) + a_2},
\end{equation}
where $a_2$ is another tunable hyper-parameter of the model to control the gap among singular values. Moreover, the authors recognize the importance of different types of relationships during the message-passing (i.e., user-item, user-user, item-item). For this reason, they decide to augment the loss function with other components addressing also the similarities among node embeddings from the same partition:
\begin{equation}
    \begin{aligned}
        \min\limits_{\mathbf{e}_v, \mathbf{e}_w, \mathbf{e}_j} \text{ }&- \sum\limits_{(v, w) \in (\mathbf{R}^{\mathcal{*}}_s)^{+}} log(sig(\mathbf{e}^\top_v \cdot \mathbf{e}_w)) \text{ }+ \\
        &- \sum\limits_{(v, j) \in (\mathbf{R}^{\mathcal{*}}_s)^{-}} log(sig(-\mathbf{e}^\top_v \cdot \mathbf{e}_j)),
    \end{aligned}
\end{equation}
where $v$, $w$, and $j$ are nodes from the same partition, and $\mathbf{R}^*$ is the interaction matrix of that partition.

\vspace{1em}
\noindent\textsc{\bfseries Observation.} \textit{The analyzed GNN models explicitly utilize the node degree information during the representation learning phase, each of them in a different way. However, clustering coefficient and degree assortativity, which share similarities with node degree's semantics, do not seem to have an evident representation within the models' formulations. Under this perspective, our study will also serve to test what topological aspects GNNs-based recommender systems can (un)intentionally capture during their training.}
\vspace{1em}

This observation paves the way to a further question: \textit{\textbf{are (topological) dataset characteristics influencing the recommendation performance of GNNs-based recommender systems?}}
\section{Proposed Evaluation Pipeline}

To answer such a question, we present our proposed evaluation pipeline to assess the impact of \textit{classical} and \textit{topological} characteristics on the performance of GNNs-based recommender systems. As already done in similar works~\cite{DBLP:journals/tmis/AdomaviciusZ12, DBLP:conf/sigir/DeldjooNSM20, DBLP:journals/corr/abs-2308-10778}, our goal is to design an explanatory statistical model which finds dependencies between dataset characteristics and recommendation performance. 

To this end, the involved pipeline steps are (i) collection of a large pool of recommendation datasets, (ii) calculation of their \textit{classical} and \textit{topological} data characteristics, (iii) train/evaluation of the GNNs-based recommender systems on the recommendation datasets to collect the performance measures (leveraging the same experimental setting as presented in~\cite{DBLP:conf/recsys/AnelliMPBSN23}), (iv) design and fit of the explanatory framework on the so-collected data characteristics/recommendation performance samples.

\subsection{Step 1: Recommendation data collection}

To fit our explanatory framework, it becomes imperative to collect a set of samples (with dataset characteristics and the corresponding models' recommendation performance) that is large enough to ensure the statistical significance of the conducted analysis under a certain confidence threshold. While in principle this process would imply the adoption of several recommendation datasets from the literature, this is in fact infeasible for time and computational reasons because it would require training and evaluating the GNNs-based models on a \textbf{\textit{very high number}} of \textbf{\textit{large}} datasets. 

To this end, and inspired by similar works~\cite{DBLP:journals/tmis/AdomaviciusZ12, DBLP:conf/sigir/DeldjooNSM20}, we propose to select some popular recommendation datasets, and manipulate them through dataset sampling strategies~\cite{DBLP:conf/sigir/WuWF0CLX21, DBLP:conf/kdd/ChenSSH17, DBLP:conf/wsdm/SachdevaWM22}  for the generation of several small recommendation datasets to conduct our study effectively. However, given the specific nature of the recommendation models we are dealing with (i.e., GNNs-based recommender systems) and differently from~\cite{DBLP:journals/tmis/AdomaviciusZ12, DBLP:conf/sigir/DeldjooNSM20}, we decide to use ad-hoc \textbf{\textit{graph}} sampling strategies such as node- and edge-dropout, which have gained recent attention in graph learning literature~\cite{DBLP:conf/sigir/WuWF0CLX21, DBLP:conf/www/ShuXLWKM22}. Indeed, as we vary the dropout rate, \textit{\textbf{we seek to collect samples that still resemble the original recommendation datasets but can encompass a wide set of topological graph structures}}. 

As for the original recommendation datasets, we use specific versions of Yelp2018~\cite{DBLP:conf/cikm/PengSM22}, Gowalla~\cite{DBLP:conf/sigir/0001DWLZ020}, and Amazon-Book~\cite{DBLP:conf/www/WangHWYL0C21}. The usage of such datasets is motivated by their popularity in GNNs-based recommendation~\cite{DBLP:conf/sigir/0001DWLZ020, DBLP:conf/cikm/MaoZXLWH21, DBLP:conf/www/LinTHZ22, DBLP:conf/cikm/GongSWLL22}. Yelp2018~\cite{DBLP:journals/corr/Asghar16} collects data about users and businesses interactions, Amazon-Book is a sub-category of the Amazon dataset~\cite{DBLP:conf/www/HeM16}, and Gowalla~\cite{DBLP:conf/kdd/ChoML11} is a social-based dataset where users share their locations.

Given the original recommendation dataset, represented as a bipartite and undirected user-item graph, we first randomly select a dropout rate in the range [0.7, 0.9] and a graph sampling strategy in \{node-dropout, edge-dropout\}. Second, the sampled graph is obtained by either dropping at random nodes or edges depending on the selected dropout rate and sampling strategy. This process is iteratively repeated for each original dataset; for the sake of this work, and following~\cite{DBLP:journals/tmis/AdomaviciusZ12, DBLP:conf/sigir/DeldjooNSM20}, we decide to collect 600 samples from each original data, for a total of 1,800 generated samples. 

\subsection{Step 2: Characteristics calculation}

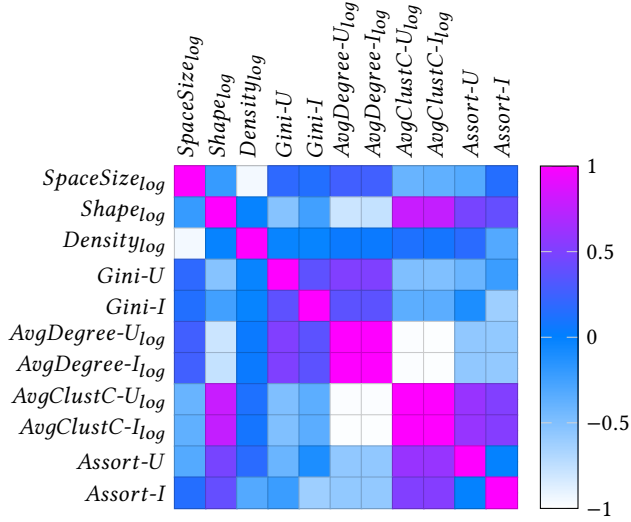
\begin{figure}[!t]
\centering
\begin{tikzpicture}
\begin{axis}[
    axis equal image, 
    scatter, 
    colormap/cool, 
    colorbar, 
    point meta min=-1,
    point meta max=1,
    grid=none, 
    minor tick num=1, 
    tickwidth=0pt, 
    y dir=reverse, 
    xticklabel pos=right, 
    xticklabel style={rotate=90},
    xtick={0,1,2,3,4,5,6,7,8,9,10},
    xticklabels={$SpaceSize_{log}$, $Shape_{log}$, $Density_{log}$, $Gini\text{-}U$, $Gini\text{-}I$, $AvgDegree\text{-}U_{log}$, $AvgDegree\text{-}I_{log}$, $AvgClustC\text{-}U_{log}$, $AvgClustC\text{-}I_{log}$, $Assort\text{-}U$, $Assort\text{-}I$},
    ytick={0,1,2,3,4,5,6,7,8,9,10},
    yticklabels={$SpaceSize_{log}$, $Shape_{log}$, $Density_{log}$, $Gini\text{-}U$, $Gini\text{-}I$, $AvgDegree\text{-}U_{log}$, $AvgDegree\text{-}I_{log}$, $AvgClustC\text{-}U_{log}$, $AvgClustC\text{-}I_{log}$, $Assort\text{-}U$, $Assort\text{-}I$},
    enlargelimits={abs=0.5}, 
    scatter/@pre marker code/.append code={
      \pgfplotstransformcoordinatex{sqrt(abs(\pgfplotspointmeta))}
      \scope[mark size=6, fill=mapped color]
    },
    scatter/@post marker code/.append code={%
      \endscope%
    },
    scale=0.8 
]
\addplot +[
    point meta=explicit, 
    mark=square*,
    only marks, 
    ] table [
    x expr={int(mod(\coordindex+0.01,11))}, 
    y expr={int((\coordindex+0.01)/11))},
    meta=value
] {
X   Y   value

0	0	1.000000
0	1	-0.212315
0	2	-0.954520
0	3	0.180507
0	4	0.134404
0	5	0.255242
0	6	0.254337
0	7	-0.408143
0	8	-0.377079
0	9	-0.328141
0	10	0.143009
1	0	-0.212315
1	1	1.000000
1	2	-0.021688
1	3	-0.524974
1	4	-0.254467
1	5	-0.794564
1	6	-0.767826
1	7	0.783785
1	8	0.765346
1	9	0.469329
1	10	0.394994
2	0	-0.954520
2	1	-0.021688
2	2	1.000000
2	3	-0.031705
2	4	-0.027120
2	5	0.044572
2	6	0.045499
2	7	0.118263
2	8	0.084351
2	9	0.163524
2	10	-0.325265
3	0	0.180507
3	1	-0.524974
3	2	-0.031705
3	3	1.000000
3	4	0.371105
3	5	0.506309
3	6	0.497189
3	7	-0.497250
3	8	-0.503362
3	9	-0.415233
3	10	-0.228037
4	0	0.134404
4	1	-0.254467
4	2	-0.027120
4	3	0.371105
4	4	1.000000
4	5	0.361340
4	6	0.363311
4	7	-0.362226
4	8	-0.355434
4	9	-0.111091
4	10	-0.620275
5	0	0.255242
5	1	-0.794564
5	2	0.044572
5	3	0.506309
5	4	0.361340
5	5	1.000000
5	6	0.999082
5	7	-0.984254
5	8	-0.989383
5	9	-0.569820
5	10	-0.573792
6	0	0.254337
6	1	-0.767826
6	2	0.045499
6	3	0.497189
6	4	0.363311
6	5	0.999082
6	6	1.000000
6	7	-0.983229
6	8	-0.989941
6	9	-0.568128
6	10	-0.577563
7	0	-0.408143
7	1	0.783785
7	2	0.118263
7	3	-0.497250
7	4	-0.362226
7	5	-0.984254
7	6	-0.983229
7	7	1.000000
7	8	0.998075
7	9	0.598835
7	10	0.511454
8	0	-0.377079
8	1	0.765346
8	2	0.084351
8	3	-0.503362
8	4	-0.355434
8	5	-0.989383
8	6	-0.989941
8	7	0.998075
8	8	1.000000
8	9	0.594505
8	10	0.523383
9	0	-0.328141
9	1	0.469329
9	2	0.163524
9	3	-0.415233
9	4	-0.111091
9	5	-0.569820
9	6	-0.568128
9	7	0.598835
9	8	0.594505
9	9	1.000000
9	10	-0.013674
10	0	0.143009
10	1	0.394994
10	2	-0.325265
10	3	-0.228037
10	4	-0.620275
10	5	-0.573792
10	6	-0.577563
10	7	0.511454
10	8	0.523383
10	9	-0.013674
10	10	1.000000
};
\end{axis}
\end{tikzpicture}
\caption{Pearson correlation of the selected characteristics. Many values in $[-0.5, 0.5]$ indicate loosely correlated pairs.}
\label{fig:correlation}
\end{figure}

After the recommendation data collection, the second pipeline step involves the calculation of \textit{classical} and \textit{topological} characteristics of the samples. While we have already focused on the selected \textit{topological} properties, in terms of \textit{classical} characteristics, we select space size, shape, density, and Gini coefficient calculated on the user and item side (once again, we suggest the reader refer to ~\cite{DBLP:journals/tmis/AdomaviciusZ12, DBLP:conf/sigir/DeldjooNSM20} for a formal presentation). Then, inspired by similar works~\cite{DBLP:journals/tmis/AdomaviciusZ12, DBLP:conf/sigir/DeldjooNSM20}, we decide to apply the log10-scale to the formulation of some characteristics to obtain values within comparable order of magnitude, thus making the training of the explanatory model more stable. In the remaining of the paper, for the sake of easy understanding, we use the following notation to indicate the selected data characteristics: $SpaceSize_{log}$, $Shape_{log}$,$ Density_{log}$,$ Gini\text{-}U$, $Gini\text{-}I$, $AvgDeg\text{-}U_{log}$, $AvgDeg\text{-}I_{log}$, $AvgClustC\text{-}U_{log}$, $AvgClustC\text{-}I_{log}$, $Asso\text{-}\\rt\text{-}U$, $Assort\text{-}I$, where ``log'' refers to the log10-scale normalization. Empirically, we note that the vast majority of calculated characteristics appear loosely correlated, hence further supporting their adoption for our analysis (\Cref{fig:correlation}).

\subsection{Step 3: Train/test GNNs-based models}

As a third pipeline step, we train and evaluate the selected GNNs-based recommender systems on the pool of generated datasets. To this end, we follow the same experimental setting proposed in~\cite{DBLP:conf/recsys/AnelliMPBSN23}, and use Elliot as the reproducibility framework to run our experiments~\cite{DBLP:conf/sigir/AnelliBFMMPDN21, DBLP:conf/um/MalitestaPANF23}. Thus, we perform the random subsampling strategy to split each dataset into train and test (80\% and 20\%, respectively). Then, we retain the 10\% of the train as validation for the early stopping to avoid overfitting. To train LightGCN, DGCF, and SVD-GCN, we fix their configurations (i.e., hyper-parameters and patience for the early stopping) to the best values according to the original papers, since our scope is not to fine-tune them. Finally, following the literature, we use the Recall@20 calculated on the validation for the early stopping and evaluate the models by assessing the same metric on the test set. To foster the reproducibility of the results, we share the GitHub repository with all codes, datasets, configuration files, and documentation: \url{https://github.com/sisinflab/Topology-Graph-Collaborative-Filtering}. 

\subsection{Step 4: Explanatory model}

Finally, we aim to fit an explanatory model to the collected dataset characteristics/recommendation performance samples. Statistical models can be utilized to elucidate the relationship between a hypothesized cause of a phenomenon (i.e., independent variables) and its effect (measured through dependent variables). While various potential functions can be used to fit the independent variables to the dependent ones, we opt to utilize a linear regression model for two reasons: (i) to adhere to the same methodology employed in recent studies such as~\cite{DBLP:journals/tmis/AdomaviciusZ12,DBLP:conf/sigir/DeldjooNSM20}, and (ii) to derive explanations on the performance impact of data characteristics through linear dependencies, which represents the most straightforward and intuitive strategy. From this intuition, we use a regression model:
\begin{equation}
\mathbf{y}=\boldsymbol{\epsilon}+\theta_0+\boldsymbol{\theta}_c \mathbf{X}_c.
\label{base_formula}
\end{equation}

We recall that our goal is to test if the factors related to the data characteristics (i.e., $\mathbf{X}_{c}$) can explain the effect on the recommendation system's performance (i.e., $\mathbf{y}$). Therefore, in Equation \ref{base_formula}, we denote by $\boldsymbol{\theta}_c=\left[\theta_1, \ldots, \theta_C\right]$ the vector of regression coefficients each of whom is associated with the $c$-th feature (data characteristic considered here), $\mathbf{X}_c \in \mathbb{R}^{M \times C}$ the matrix containing the data characteristic values for each sample in the training set, and $\mathbf{y}$ the vector containing the values of the performance measure associated with all samples in the training set. Moreover, under the assumption of mean-centered data, $\theta_0$ expresses the expected value of $\mathbf{y}$ (i.e., in this case, the expected recommendation performance). The regression model is trained through Ordinary Least Squares (OLS):
\begin{equation}
\label{eq:OLS}
\left(\theta_0^*, \boldsymbol{\theta}_c^*\right)=\min _{\theta_0, \boldsymbol{\theta}_c} \frac{1}{2}\left\|\mathbf{y}-\theta_0-\boldsymbol{\theta}_c    \mathbf{X}_c\right\|_2^2.
\end{equation}

To show how the recommendation performance is related to dataset characteristics, we utilize the basic regression model presented in Equation~\ref{eq:OLS} to maximize the $R^2$ coefficient. This approach allows us to effectively motivate the impact of the $\boldsymbol{\theta}_{c}$ coefficients on the recommendation system's effectiveness, as outlined in~\cite{gareth2013introduction} for any regression model.

\section{Results and Discussion}

We aim to answer two research questions: \textbf{RQ1)} What is the impact of \textit{classical} and \textit{topological} characteristics on the performance of GNNs-based recommender systems?; \textbf{RQ2)} Is the dataset generation through node- and edge-dropout differently influencing the explanations of our model? 

\subsection{Impact of characteristics (RQ1)}
\begin{figure*}[!t]
\centering

\subfloat[LightGCN]{
    \includegraphics[width=0.85\textwidth]{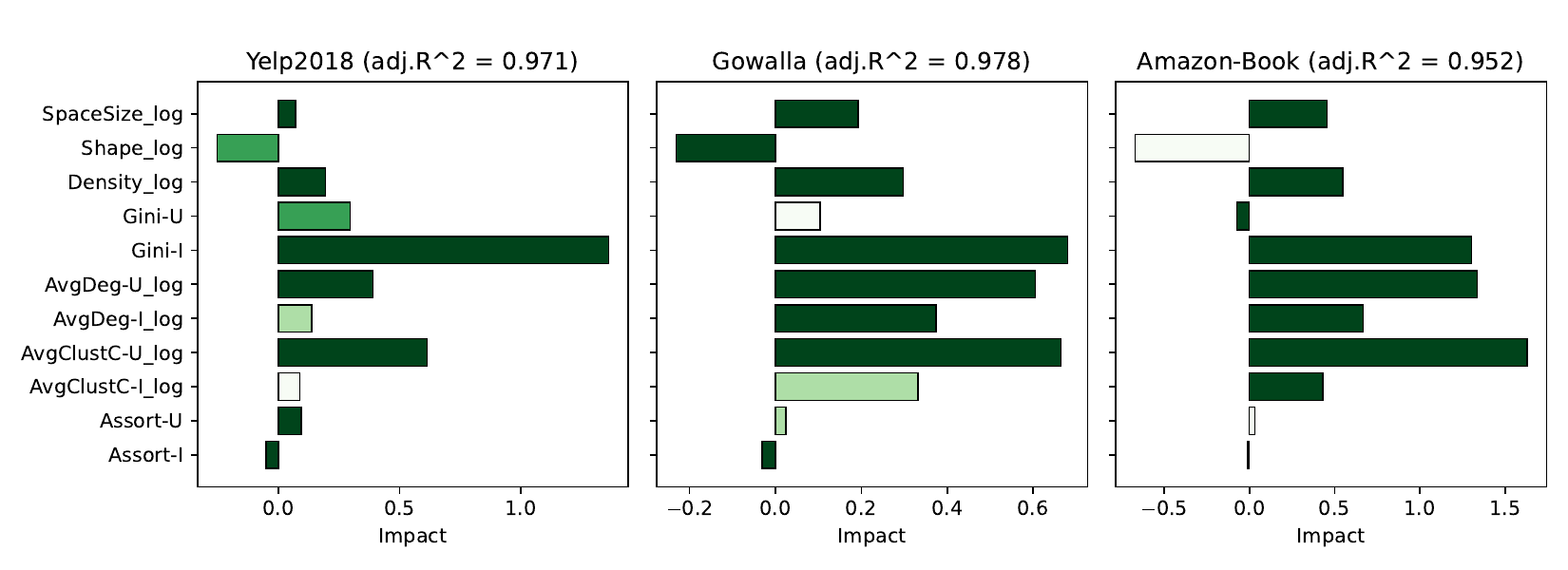}}

\vspace{-1em}

\subfloat[DGCF]{
    \includegraphics[width=0.85\textwidth]{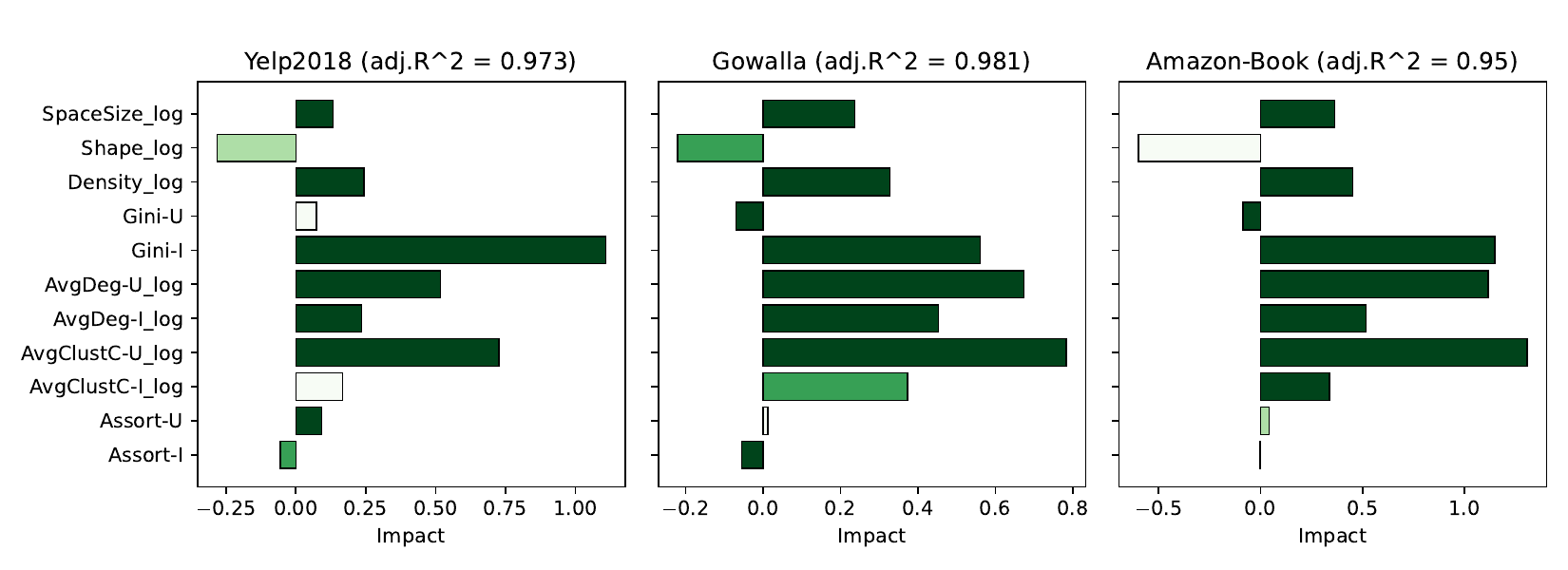}}

\vspace{-1em}

\subfloat[SVD-GCN]{
    \includegraphics[width=0.85\textwidth]{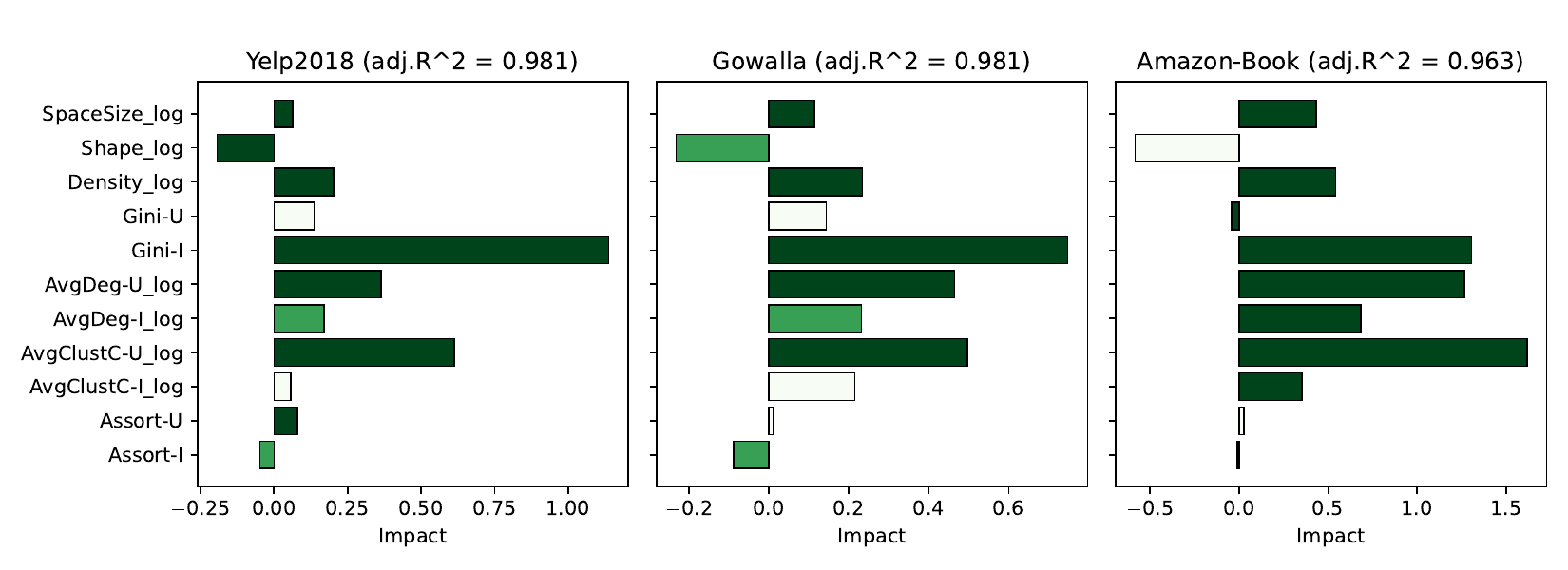}}

\vspace{-1em}

\subfloat{
    \includegraphics[width=0.65\textwidth]{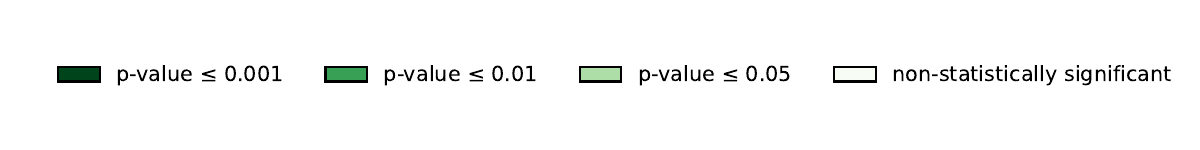}}

\vspace{-1em}

\caption{Visual representation of the impact of dataset characteristics on the recommendation performance (Recall@20) of GNNs-based recommender systems, for each dataset/model setting. Bar plot length and direction represent the impact magnitude and whether there is a direct/inverse correspondence between characteristic and performance. Finally, the darker the bar plots, the higher their statistical significance.}\label{fig:rq1}
\end{figure*}

We assess the impact of \textit{classical} and \textit{topological} characteristics on the accuracy performance (i.e., Recall@20) of GNNs-based recommender systems. \Cref{fig:rq1} displays, for each dataset/GNNs-based model setting, the learned regression coefficients of the explanatory framework. Specifically, each bar plot indicates the relative impact of a given data characteristic on the Recall@20 performance; moreover, bar plot length and direction represent the magnitude of this impact and whether there exists a direct/inverse correspondence between characteristic and performance. Finally, to assess the goodness of the results, we also estimate the adj. (adjusted) $R^2$~\cite{DBLP:journals/tmis/AdomaviciusZ12,DBLP:conf/sigir/DeldjooNSM20} of the regression model, along with the statistical significance of the learned coefficients (the darker the bar plots, the higher the statistical significance). Overall, the adj. $R^2$ is above 95\%, proving the ability of the regression model to explain the accuracy recommendation performance through the measured characteristics. Hereinafter, we further decompose the regression results by categorizing the characteristics into \textit{classical} and \textit{topological}.

\subsubsection{Classical characteristics} Previous works~\cite{DBLP:journals/tmis/AdomaviciusZ12} have assessed the impact of \textit{classical} characteristics on neighbor- and factorization-based models separately. However, a careful search of the relevant literature yields that no study has investigated whether such characteristics influence GNNs-based recommender systems~\cite{DBLP:journals/corr/abs-2308-10778}; as already discussed, GNNs-based models involve both a neighborhood- and factorization-based module, where the former corresponds to the message-passing, and the latter to the users' and items' latent factors as in traditional latent-based recommendation~\cite{DBLP:conf/uai/RendleFGS09, DBLP:conf/icdm/Rendle10}.

Interestingly, \Cref{fig:rq1} suggests that, in GNNs-based recommender systems, the factorization component might be more influential than the neighborhood one on the recommendation performance. In this respect, if we refer to the results from~\cite{DBLP:journals/tmis/AdomaviciusZ12}, we observe that factorization- and GNNs-based approaches are particularly aligned, considering: (i) the inverse correspondence between the performance metric and the $Shape_{log}$ in almost all settings, meaning that when the number of users is higher than the number of items in the system, performance may decrease; (ii) the direct correspondence between the performance metric and $Density_{log}$ and $Gini\text{-}I$. As for (ii), the density is historically known as one of the core problems in recommendation (i.e., data sparsity), so it becomes evident why also GNNs-based recommender systems' performance benefits from denser (i.e., less sparse) datasets. The Gini index measures the dissimilar distribution of items' interactions in the system and could be related to the tendency of recommender systems to promote popular items from the catalog. That is, when there exist items that have been experienced more frequently than others, both GNNs- and factorization-based recommender systems may be biased towards popular items, and so their accuracy performance increases. Noteworthy, all such observations are supported by the statistical significance of the results.

\subsubsection{Topological characteristics} GNNs-based models interpret the user-item data as a graph. Consequently, our novel evaluation analyzes the influence of \textit{topological} characteristics of the selected recommendation datasets on the recommendation performance.

The most evident outcome is that $AvgDegree\text{-}U_{log}$ and $AvgDe\-gree\text{-}I_{log}$ show a direct correspondence with performance in almost all settings. Indeed, this analytically confirms the intuitions provided in the literature on the recommendation scenario~\cite{DBLP:conf/recsys/AnelliMPBSN23} or for other tasks~\cite{DBLP:journals/corr/abs-2310-04612}, as well as what we already observed in \Cref{sec:top-char-graph-collab} regarding the explicit presence of the node degree in the formulations of all the selected GNNs-based approaches. In practical terms, when GNNs-based models are trained on datasets with several interactions for users and items, they learn accurate users' preferences since each node receives the contribution of numerous neighbor nodes. It is worth noticing that, in absolute values, $AvgDegree\text{-}U_{log}$ is more influential than $AvgDegree\text{-}I_{log}$ on the overall performance. Hence, under the same average degree gain, a user average degree improvement is preferable since it would lead to better performance.

As far as clustering coefficient and degree assortativity are concerned, we assess how similarities among nodes from the same partition in the graph may impact the recommendation accuracy performance of models. In terms of $AvgClustC\text{-}U_{log}$ and $AvgClustC\text{-}I_{log}$, the results prove again a strong direct correspondence in almost all settings of GNNs-based models and datasets. Differently from the average degree scenario, the relative importance of the user-side values is much higher than the one of the item-side for LightGCN and DGCF, while the gap sometimes gets narrower in the case of SVD-GCN. This may happen because while LightGCN and DGCF only leverage user-item types of interactions, SVD-GCN also embed the information conveyed in the user- and item-projected graphs in its formulation, thus flattening the different influence of the user-side characteristics over the item-side counterpart. 

Interestingly, the $Assort\text{-}U$ and $Assort\text{-}I$ characteristics exhibit a direct and inverse correspondence to the recommendation metric, respectively. Furthermore, models such as LightGCN and DGCF have slightly larger coefficients for both $Assort\text{-}U$ and $Assort\text{-}I$ than SVD-GCN. Again, these results have a mathematical justification. Indeed, the strong \textit{lookahead} nature of the assortativity measures (refer again to \Cref{sec:topological_char}) seems to be captured by the multi-layer message-passing performed by LightGCN and DGCF.
Conversely, in the case of SVD-GCN, they are less influential, probably because the model acts on the singular values of the adjacency matrix with the effect of limiting the graph convolutional layers' depth to avoid over-smoothing.
However, the assortativity results are less significant than the others, so we plan to further investigate this aspect. 

\vspace{1em}
\noindent\textsc{\bfseries Summary.} \textit{The results show that: (i) interestingly, the factorization component of GNNs-based recommender systems seems to be impacting the recommendation performance more than the neighborhood (message-passing) component; (ii) while confirming its influence on the recommendation performance, node degree seems not to be a key topological characteristic to distinguish among the different GNNs-based models; indeed, the wider perspective provided by clustering coefficient and (especially) degree assortativity may help to recognize how the different models address the topological properties of the graph, even with unexpected outcomes.}
\vspace{1em}

\subsection{Influence of node- and edge-dropout (RQ2)} 
\label{sec:rq2}
The current section investigates the influence of node- and edge-dropout on the explanatory model. In the interest of space, we report an extensive analysis of the largest dataset, Gowalla, by considering the performance of SVD-GCN, the most recent GNNs-based approach among the selected ones. To answer the RQ, we provide both a \textit{theoretical} and \textit{analytical} intuition.

\subsubsection{Theoretical intuition} \Cref{fig:scale-free} displays the relation between the probability distribution of node degrees in the original graph and their degree values on the Gowalla dataset. As evident, high-degree nodes are less popular than low-degree ones, and this resembles the tendency of real-world networks to be \textit{scale-free}~\cite{PhysRevE.67.026112}. To be more precise, the actual degree probability distribution approximates neither the \textit{power-law} (i.e., representing \textit{scale-free} networks, in green), nor the \textit{exponential} function (i.e., in red), but it would be approximated by a function in-between. Thus, high-degree nodes are even less frequent than they usually are in \textit{scale-free} networks.

The figure helps us understand what the different impacts of node- and edge-dropout on such a graph topology might be. By recalling that node-dropout works by removing nodes (and all the edges connected to them) and edge-dropout eliminates edges and the consequently disconnected nodes, in the worst-case scenario, node-dropout would remove many high-degree nodes from the graph; instead, edge-dropout would eliminate all the edges connected to several nodes and thus disconnect them from the graph. Hence, on average, node-dropout would drop larger portions of the user-item graph than edge-dropout, possibly undermining the goodness of the explanations produced by our explanatory framework. This intuition \textit{theoretically} justifies the \textit{\textbf{joint}} adoption of node- and edge-dropout for the experiments presented in RQ1.

\input{figures/scale_free}

\subsubsection{Analytical intuition} To further analytically test the previous intuition, we build four versions of the dataset $\mathbf{X}_c$ for the explanatory framework, each obtained with varying portions of node- and edge-dropout, respectively (refer again to \Cref{base_formula}). Specifically, the number of samples in $\mathbf{X}_c$ changes in accordance to:
\begin{equation}
    |\mathbf{X}_c| = (1 - \alpha) |\mathbf{X}^{n}_{c}| + \alpha |\mathbf{X}^{e}_{c}|,
\end{equation}
where $\mathbf{X}^{n}_{c}$ and $\mathbf{X}^{e}_{c}$ indicate the portion of $\mathbf{X}_c$ sampled through node- and edge-dropout, while $\alpha$ is a parameter to control the number of samples from $\mathbf{X}^{n}_{c}$ and $\mathbf{X}^{e}_{c}$ contributing to the final dataset $\mathbf{X}_c$. We use $|\cdot|$ as a necessary notation abuse to refer to any dataset size in a simple way. We let $\alpha$ range in $\{0.0, 0.3, 0.7, 1.0\}$, where extreme values of $\alpha$ are used to build the dataset through either node- or edge-dropout; the others combine the two sampling strategies.

\begin{table*}[!t]
\caption{Explanatory framework results with varying values of $\alpha$ on Gowalla for SVD-GCN. In the first row, we provide a visual intuition for the $\alpha$ value changes in all four settings. In the second row, we report the average sampling statistics of the datasets. Finally, in the third row, we indicate the adj.$R^2$ regression measure, and a pie chart indicating, for each learned regression coefficient, its statistical significance (the darker the more statistically significant).}
\label{tab:rq2}
    \centering
    \large
    \begin{tabular}{c|c|c|c}
    \hline
        \textbf{Node} \tikzcircle[fill=black]{2.5pt} \tikzcircle[fill=black]{2.5pt} \tikzcircle[fill=black]{2.5pt} \quad
         \textbf{Edge} \tikzcircle[fill=white]{2.5pt} \tikzcircle[fill=white]{2.5pt} \tikzcircle[fill=white]{2.5pt} & \textbf{Node} \tikzcircle[fill=black]{2.5pt} \tikzcircle[fill=black]{2.5pt} \tikzcircle[fill=white]{2.5pt} \quad
         \textbf{Edge} \tikzcircle[fill=black]{2.5pt} \tikzcircle[fill=white]{2.5pt} \tikzcircle[fill=white]{2.5pt} & \textbf{Node} \tikzcircle[fill=black]{2.5pt} \tikzcircle[fill=white]{2.5pt} \tikzcircle[fill=white]{2.5pt} \quad
         \textbf{Edge} \tikzcircle[fill=black]{2.5pt} \tikzcircle[fill=black]{2.5pt} \tikzcircle[fill=white]{2.5pt} & \textbf{Node} \tikzcircle[fill=white]{2.5pt} \tikzcircle[fill=white]{2.5pt} \tikzcircle[fill=white]{2.5pt} \quad
         \textbf{Edge} \tikzcircle[fill=black]{2.5pt} \tikzcircle[fill=black]{2.5pt} \tikzcircle[fill=black]{2.5pt} \\ \hline
         
        \textit{\ul{Avg. Stats.}} & \textit{\ul{Avg. Stats.}} & \textit{\ul{Avg. Stats.}} & \textit{\ul{Avg. Stats.}} \\ 
        \textbf{Users:} 5,828 & \textbf{Users:} 12,744 & \textbf{Users:} 21,730 & \textbf{Users:} 28,526 \\ 
        \textbf{Items:} 7,887 & \textbf{Items:} 17,229 & 
        \textbf{Items:} 29,316 & 
        \textbf{Items:} 38,467 \\ \textbf{Interactions:} 45,620 & \textbf{Interactions:} 97,785 & \textbf{Interactions:} 160,919 & \textbf{Interactions:} 209,659 \\\hline
        
        \multicolumn{1}{c|}{} & \multicolumn{1}{c|}{} & \multicolumn{1}{c|}{} \\ 
        \multicolumn{1}{c|}{adj.$R^2$ = 0.745} & \multicolumn{1}{c|}{adj.$R^2$ = 0.969} & \multicolumn{1}{c|}{adj.$R^2$ = 0.987} & adj.$R^2$ = 0.991 \\ \multicolumn{1}{c|}{} & \multicolumn{1}{c|}{} & \multicolumn{1}{c|}{} \\
        
        \multicolumn{1}{c|}{\includegraphics[width=0.1\textwidth]{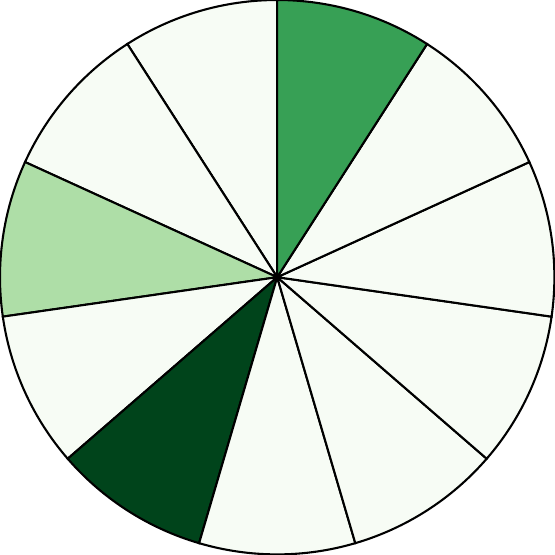}} & 
        \multicolumn{1}{c|}{\includegraphics[width=0.1\textwidth]{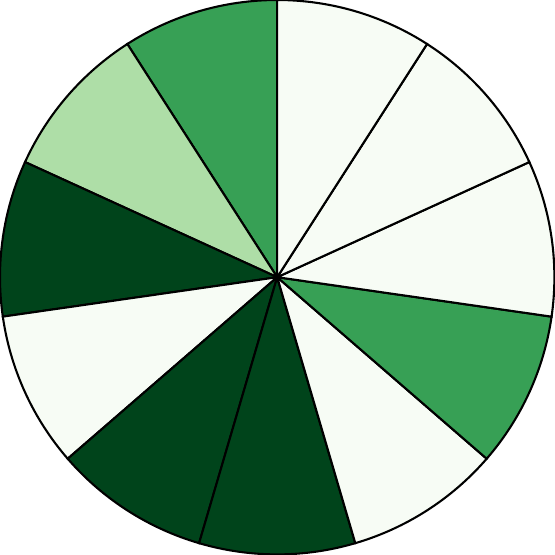}} & \multicolumn{1}{c|}{\includegraphics[width=0.1\textwidth]{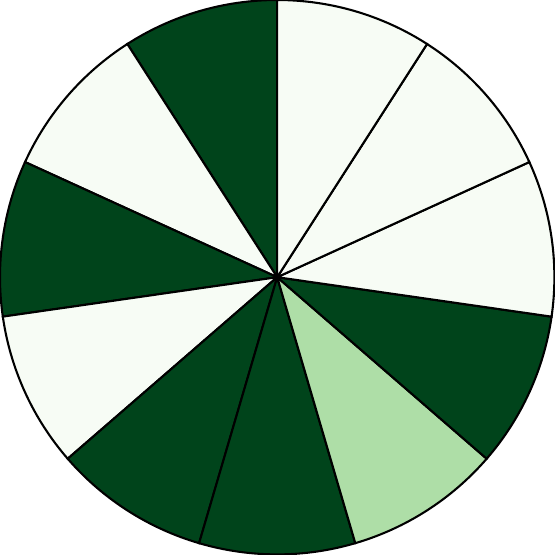}} & \includegraphics[width=0.1\textwidth]{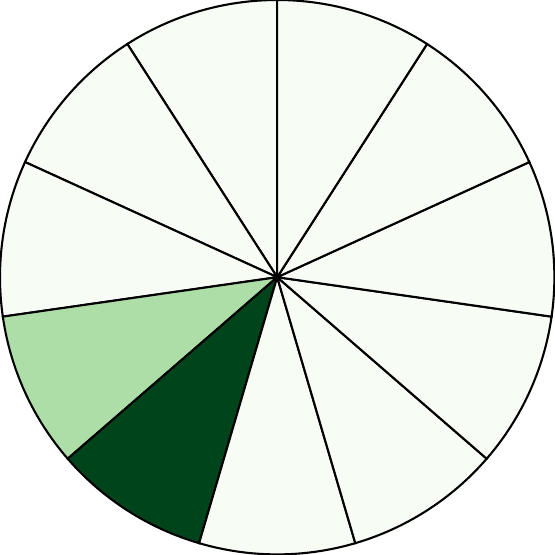}
        \\ \multicolumn{1}{c|}{} & \multicolumn{1}{c|}{} & \multicolumn{1}{c|}{} \\
    \hline
    \multicolumn{4}{c}{\includegraphics[width=0.6\textwidth]{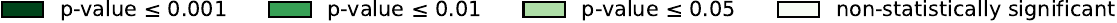}}\\
    \hline
    \end{tabular}
\end{table*}

\Cref{tab:rq2} provides a visual representation of how results vary with different $\alpha$ values. Note that we only report the average sampling statistics and the measures indicating the goodness of the generated explanations (i.e., the calculated adj.$R^2$, and statistical significance of the learned regression coefficients represented in shades of green once again) as this is the only information we need to answer the RQ. In alignment with the \textit{theoretical} intuition, the average sampling statistics show that node-dropout generally retains smaller portions of the graph than the edge-dropout. Then, the regression results highlight that the optimal trade-off between high adj. $R^2$ and statistical significance of the learned coefficients is reached when combining samples generated through both node- and edge-dropout. On the contrary, the settings with either node- or edge-dropout do not offer the conditions for the regression model to learn meaningful dependencies. This \textit{analytically} justifies the \textit{\textbf{joint}} adoption of node- and edge-dropout for RQ1.

\vspace{1em}
\noindent\textsc{\bfseries Summary.} \textit{The theoretical and analytical evaluation of the explanatory model for different settings of node- and edge-dropout indicates that their joint combination (i.e., the strategy we followed in RQ1) is beneficial to produce meaningful explanations.}
\section{Conclusion and Future Work}

In this work, we propose a novel evaluation perspective on graph neural networks (GNNs)-based recommender systems, that aims to assess the influence of the user-item graph topological properties on the recommendation performance of such models. First, we select classical and topological properties of the recommendation data, as well as three popular and recent GNNs-based recommender systems. On such a basis, we design a novel evaluation pipeline, involving: (i) the generation of a large pool of reduced-size recommendation data (sampled from the selected datasets through node- and edge-dropout) that encompass a wide range of topological structures; (ii) the calculation of their dataset characteristics and (iii) evaluation of the models' recommendation performance; (iv) the design and fit of an explanatory model that finds linear dependencies between dataset characteristics and performance. Results, validated by statistical tests and under different sampling settings, largely demonstrate the presence of strong characteristics/performance correspondences, offering a novel perspective on GNNs-based recommendation. We plan to extend our investigation to other GNNs-based recommender systems and consider additional recommendation metrics (e.g., bias and fairness dimensions). Moreover, driven by the derived insights, we aim to propose a novel GNNs-based approach that exploits the topological structure of the graph to eventually boost the recommendation performance.

\begin{acks}
The authors acknowledge the partial support by: CT\_FINCONS\_III, OVS Fashion Retail Reloaded, LUTECH DIGITALE 4.0, Secure Safe Apulia, IDENTITA, REACH-XY. The authors acknowledge the CINECA award under the ISCRA initiative for the availability of high-performance computing resources.
This work has been carried out while A. C. M. Mancino was enrolled in the Italian National Doctorate on Artificial Intelligence run by the Sapienza University of Rome in collaboration with the Polytechnic University of Bari.
\end{acks}

\bibliographystyle{ACM-Reference-Format}
\bibliography{references}

\end{document}